\documentstyle[12pt]{article}
\begin{document}
\vspace*{-2mm}
\thispagestyle{empty}
\noindent
\begin{center}
  \begin{Large}
  \begin{bf}
 Z PLUS FOUR JETS PRODUCTION IN HADRONIC COLLISIONS \\
  \end{bf}
  \end{Large}
  \vspace{0.8cm}
V. Barger and E. Mirkes\footnote{Presented by E. Mirkes.}\\[2mm]
{\it Physics Department, University of Wisconsin, Madison,
WI 53706, USA }\\
\vspace{0.3cm}
\vspace*{0.3cm}
R.J.N.~Phillips\\[2mm]
{\it Rutherford Appleton Laboratory, Chilton, Didcot, Oxon OX11 0QX, UK
     }\\
\vspace{0.3cm}
and\\
\vspace*{0.3cm}
T.~Stelzer\\[2mm]
{\it Physics Department, Durham University, Durham DH1 3LE, UK.
     }\\[2cm]
  {\bf Abstract}
\end{center}
\begin{quotation}
\noindent
We present first results of $Z+4$ jet cross sections
at the Tevatron $p\bar p$
collider  with heavy quark flavor identification.
The $Z + 4$ jet channel  is of particular interest as a normalizer for the
$W +4$ jet background to top quark signals, as a background to a possible
$t\to cZ$ flavor-changing neutral-current (FCNC) decay signal, and as a
background to missing-$p_T$ signals from gluino pairs.
We also calculate the different heavy flavor contributions
to $W+4$ jet production.
The MADGRAPH program is used to generate all leading order
subprocess helicity amplitudes. We  present Monte Carlo studies
with  separation and acceptance criteria
suitable for the Tevatron experimental analyses.
\end{quotation}

There are many potential new physics processes at hadron colliders, that
would lead to final states with a weak boson
plus multi-jets, where the weak boson is identified by its
leptonic decay; these signals sometimes also contain a
second weak boson, whose hadronic decay is less easily identified.
Since a weak boson can also be produced along with gluon and quark jets,
a knowledge of these QCD backgrounds is essential to the identification
of new physics signals.  Considerable effort has been devoted in recent years
to the calculation of QCD $W + n$ jet ($n=1,2,3,4$) and $Z + n$ jet
($n=1,2,3$) cross sections; for the cases of high
jet multiplicities $n$, that would be given by many
interesting new physics signals, these calculations
can currently be made at tree level only \cite{vecbos,hag,wisc,mangano}.
We present first results for $Z$ production with
four QCD jets, evaluated for the Tevatron $p\bar p$ collider at
$\sqrt s=1.8$~TeV, including a separation of contributions from
different heavy quark flavors.  We also calculate $W + 4$ jets
with heavy quark flavor identification; this goes beyond previous
$W + 4$ jet results that flag only $b$-flavor\cite{vecbos}.

Major areas of physics interest in a QCD $Z + 4$ jet calculation
are the following.

\noindent
(a) The most immediate interest is in the top-quark searches at the
Tevatron\cite{cdf,d0};  the single-lepton signals have
QCD $W+4$ jet production as the major background, and
the $W/Z$ ratio could provide a calibration.
We calculate separate cross sections for different final-state jet
flavors, to address the case of $b$-tagging.

\noindent
(b) Possible isosinglet heavy quarks $x$ would have both charged-current
and neutral-current decay modes, with branching fraction
ratios\cite{bargp}
\begin{equation}
B(x\to qW)\; :\; B(x\to q'Z) \simeq 2\; : \; 1 .
\end{equation}

\noindent
(c) A related question is the possible existence of a prominent FCNC
decay mode of the top quark\cite{hall}, $t\to cZ$ along with the standard
$t\to bW$ decay.  This scenario leads to a
$t\bar t\to (cZ)(bW) \to Z + 4$ jet signal, to be distinguished
from QCD background.

\noindent
(d) Supersymmetric particle production gives missing-$p_T$
plus multijet signals at hadron colliders.  In particular,
gluino pairs $\tilde g \tilde g$  with decays $\tilde g \to \chi_1^0
q \bar q$ to the lightest neutralino $\chi_1^0$ are expected to give
missing-$p_T$ plus 4 jets.    Here $Z + 4$ jet production
with $Z\to\nu\bar\nu$ is the major standard physics
background; $b$-tagging is relevant, since
there are regions of parameter space where $\tilde g\to t\tilde t$
or $\tilde g\to b\tilde b$  decays are dominant \cite{baer}.


In our $Z + 4$-jet and $W + 4$-jet calculations, many Feynman diagrams must
be expressed as amplitudes; e.g. $gg \to Zq\bar q gg$ involves 516 diagrams.
This phase is accomplished by the MADGRAPH program\cite{madgraph}, which
generates all Feynman graphs and their helicity amplitudes,
employing the HELAS approach\cite{helas}.  We then fold in final phase space
and initial parton distributions, using MRS set $D_{-}^{\prime}$\cite{mrs}
evaluated at a scale $Q^2= \left<p_T\right>^2+M_Z^2(M_W^2)$.
For semi-realistic simulations, we make parton-level calculations of
$p\bar p\to W(Z)+4$ jets at $\sqrt s = 1.8$ TeV, treating final
partons as jets if
\begin{equation}
p_T(j) > 20\,\mbox{GeV},\,\, |\eta(j)| < 2, \,\,\Delta R(jj) > 0.4,
\end{equation}
where $[\Delta R(jj)]^2 = [\Delta\eta (jj)]^2 + [\Delta\phi(jj)]^2$
defines the angular separation of jets.  A correction must be
made in comparing the parton transverse momentum $p_T$ with the
observed (uncorrected) jet transverse energy $E_T$; according to CDF
simulations\cite{cdf}, typically 5 GeV or more must be added to the latter.
The probability of $b$-tagging any particular final state
depends on the separate probabilities $\epsilon_j$
that any single jet $j=b,c,q/g$  satisfies the tagging criteria;
we shall assume $\epsilon_b=0.18$ ,
$\epsilon_c=0.05$ and $\epsilon_{q/g}=0.01$, which approximate
conditions in the CDF top-quark search\cite{cdf}.

When $Z$ is detected by $Z\to e\bar e$ and $W$ is detected
by $W\to e\nu$, we take the electron and missing transverse momentum
$p\llap/_T^{}$ acceptance to be

\begin{equation}
 p\llap/_T^{} > 20\hbox{ GeV\quad (for $W$ events)},\quad
 p_T(e) > 20\rm\ GeV, \quad |\eta(e)| < 1,
\end{equation}
and require that electrons are isolated from jets
by $\Delta R(ej) > 0.4$.  These criteria approximate those
used in Tevatron experimental analyses.
In the following $Z$ denotes $Z \to e^+ e^-$ and $W$ denotes
$W^{\pm}\to e^{\pm} \nu$ ; with these leptonic branching fractions included,
the total cross sections times branching fractions are denoted $B\sigma$,
and we obtain $B\sigma(Z + 4 \mbox{jet}) = 20.5$ fb,
$B\sigma(W + 4 \mbox{jet}) = 337$ fb .  Separate cross sections in fb for the
most important final jet-flavor configurations are:
\begin{equation}
\begin{array}{cccll}
b  & c & q/g & B\sigma (Z+4\,\mbox{jets}) & B\sigma (W+4\,\mbox{jets}) \\
2  & - &  2  & 1.03               &   10.2            \\
1  & - &  3  & 0.14               &   0.65            \\
-  & 2 &  2  & 0.88               &   10.2            \\
-  & 1 &  3  & 0.24               &   18.6            \\
-  & - &  4  & 18.1               &   297             \\
\end{array}
\end{equation}

 Folding in $b$-tagging efficiencies, the tagged cross sections are:
\begin{equation}
\begin{array}{ccc}
\mbox{no. of tags}  &  B\sigma (Z+4\,\mbox{jets}) & B\sigma (W+4\,\mbox{jets})
\\
 \geq  0       &    20.5          &      337.           \\
 \geq  1       &    1.23          &      18.1           \\
 \geq  2       &    0.057         &      0.68           \\
 \geq  3       &    0.001         &      0.01
 \end{array}
\end{equation}
To compare with
experimental event rates, one has to multiply these cross sections by
efficiency factors for the electrons and muons and also take into account
effects of detector simulations. However, these effects (together
with theoretical uncertainties) are expected to cancel approximately
in the ratio of $(W+4 \mbox{jet})/(Z+4 \mbox{jet})$ cross sections.

The predicted $W/Z$ ratio in 4-jet events with at least one $b$-tag is
about 14.7. This number is fairly insensitive to the jet threshold $p_T$
cut. Even if we relax the $p_T$ and $\eta$ requirements on the fourth
jet (as CDF does to increase statistics in the top-quark sample), this
$W/Z$ ratio remains about 14.

With 19.2~pb$^{-1}$ luminosity, CDF finds two $b$-tagged $Z+4$ jet events
and seven $b$-tagged $W+4$ jet events, with relaxed
$E_T$ and $\eta$ requirements on the fourth jet.   Although the
statistics are small, this observed  $W/Z$ ratio in 4-jet events
appears to be anomalously low in comparison with the QCD prediction.

A more detailed discussion is presented in [13].

\bibliographystyle{unsrt}

\begin{thebibliography}{99.}

\bibitem{vecbos} F.~A.~Berends et al, Nucl.Phys. {\bf B321}, 39 (1989),
 {\bf B357}, 32 (1991).

\bibitem{hag} K.~Hagiwara and D.~Zeppenfeld,  Nucl.Phys. {\bf B313},
 560 (1989).

\bibitem{wisc} V.~Barger, T.~Han, J.~Ohnemus and D.~Zeppenfeld,
 Phys. Rev. {\bf D40}, 2888  (1989).

\bibitem{mangano} Z.~Kunszt,  Nucl.Phys. {\bf B247}, 339 (1984);
 M.~Mangano,  ibid {\bf B405}, 536 (1993).

\bibitem{cdf}  CDF collaboration: F.~Abe et al, FERMILAB-PUB-94/097-E.

\bibitem{d0}   D0 collaboration: S.~Abachi et al, Phys. Rev. Lett.
 {\bf 72}, 2138 (1994).

\bibitem{bargp} For references see V.~Barger and R.J.N.~ Phillips,
 MAD/PHYS/830 (1994).

\bibitem{hall}R.~Barbieri and L.J.~Hall, Nucl. Phys.{\bf B319}, 1 (1989).

\bibitem{baer} H.~Baer, M.~Drees, C.~Kao, M.~Nojiri and X.~Tata,
 FSU-HEP-940311.

\bibitem{madgraph} T.~Stelzer and W.~F.~Long,
                   Comp.~Phys.~Comm.~{\bf 81}, 357 (1994).

\bibitem{helas} E.~Murayama, I.~Watanabe and K.~Hagiwara,
 KEK Report 91-11, January 1992.

\bibitem{mrs}
A.~D.~Martin, R.~G.~Roberts, and W.~J.~Stirling,
Phy.Lett.B {\bf 306}, 145 (1993);

\bibitem{letter} V.~Barger, R.~J.~N.~Phillips, E.~Mirkes and T.~Stelzer,
                 MAD/PH/844, July 1994, to appear in Phys.Lett.B.
\end{thebibliography}

\end{document}